# Ion Source Development For The Proposed FNAL 750keV Injector Upgrade

D.S. Bollinger[a]

[a]Fermi National Accelerator Laboratory, Box 500, Batavia, IL 60543 U.S.A.

**Abstract.**
Currently there is a Proposed FNAL 750keV Injector Upgrade for the replacement of the 40 year old Fermi National Laboratory (FNAL) Cockcroft-Walton accelerators with a new ion source and 200MHz Radio Frequency Quadruple (RFQ).[1] The slit type magnetron being used now will be replaced with a round aperture magnetron similar to the one used at Brookhaven National Lab (BNL). Operational experience from BNL has shown that this type of source is more reliable with a longer lifetime due to better power efficiency [2]. The current source development effort is to produce a reliable source with >60mA of H- beam current, 15Hz rep-rate, 100s pulse width, and a duty factor of 0.15%. The source will be based on the BNL design along with development done at FNAL for the High Intensity Neutrino Source (HINS) [3].

## INTRODUCTION

The Fermilab preinjector system consists of two Cockcroft-Walton accelerators that were installed in 1968. They have been a reliable source of H- ions since 1978 [4]. The reliability stems from having redundancy with the two Cockcroft-Walton accelerators and the gifted technicians that maintain them. However, with the retirement of critical personnel and aging parts that are no longer produced, the possibility of significant downtime has increased. To address these problems, there is a proposed plan to replace the existing preinjectors with a round aperture magnetron, RFQ and low energy beam transport (LEBT) similar to BNL. Another benefit to the BNL design is the beam quality entering the Drift Tube Linac (DTL) is much higher. The BNL emittance out of the RFQ is $\varepsilon(n,rms) \sim 0.4\pi$ mm mrad [2] round beam, whereas the FNAL Linac emiitance at the entrance to tank one of the DTL is vertical $\varepsilon(n,rms) \sim 0.83\pi$ mm mrad and horizontal $\varepsilon(n,rms) \sim 0.40\pi$ mm mrad.

# CURRENT OPERATIONS

The current operational sources are surface plasma magnetrons that have a slit aperture. The sources are mounted so that the aperture points down with a 90 degree bend magnet that helps sweep away electrons and shape the beam for injection into the accelerating column (Figure 1).

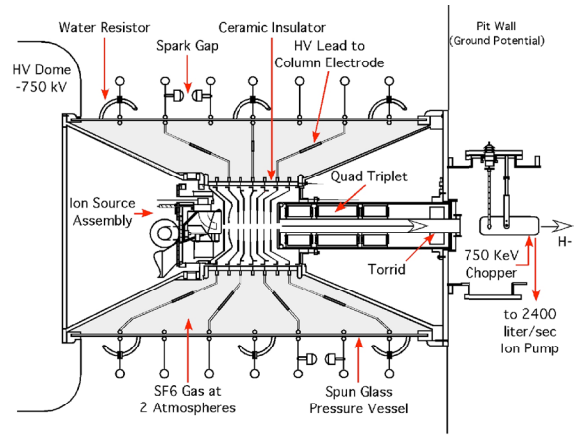

**FIGURE 1.** H- ion source Cockcroft-Walton assembly. (Image from FNAL Linac rookie book)

The H- ions are extracted through a slit opening in the anode cover plate by an H shaped extractor electrode with a positive potential of 12kV to 20kV. The extraction scheme is shown in Figure 2.

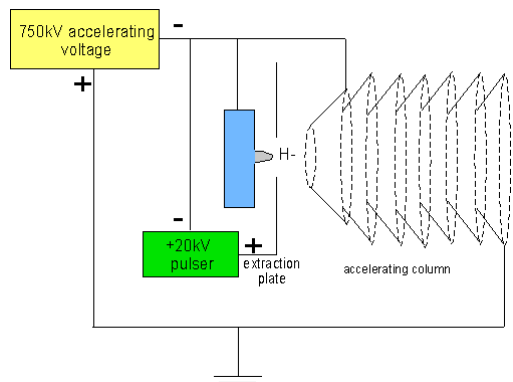

**FIGURE 2.** Schematic of the Cockcroft-Walton extraction.

The low extraction voltage requires the source to run with a high arc current to achieve the required H- beam current (table 1). With the high arc current and voltage the efficiency is on the order of 20mA/kW. The high arc current and low efficiency contribute to a source lifetime of 3 to 4 months. Typical aging of sources is caused by cathode erosion and cesium hydride blocking the hydrogen inlet aperture. Once a source is removed from operation, it is cleaned and worn parts replaced.

TABLE 1. Current ion source operational parameters

| Parameter | Value |
|---|---|
| H- current | 45-55mA |
| Arc current | 40-50A |
| Arc Voltage | 120-140V |
| Extraction Voltage | 15-18kV |
| Rep Rate | 15Hz |
| Duty Factor | 0.12% |
| Pulse Width | 80µs |
| Power efficiency | 20mA/kW |

# THE PLAN

The 750keV linac injector upgrade plan [1] for replacing the existing Cockcroft-Walton accelerators with a 200MHz RFQ and two round aperture direct magnetron ion sources as shown in Figure 3. It was decided to have two ion sources for redundancy that will be mounted on a rail system which allows one source to slide into the operational position, connected to the first solenoid, while the other has maintenance performed on it, off to the side. This allows the LEBT to be as short as possible which empirical evidence from BNL shows is desirable.

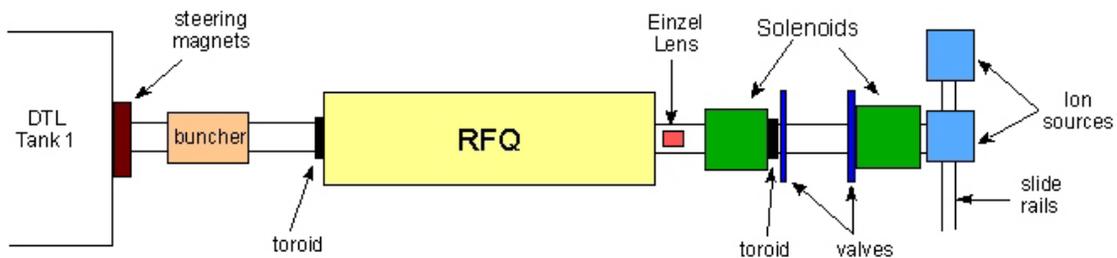

**FIGURE 3.** Proposed preinjector design.

## Magnetron Source

The round aperture magnetron has been used quite successfully at BNL since 1989 [2], shown in Figure 4, consists of a dimpled cathode that spherically focuses the H- and electrons at the anode circular aperture. The ions are extracted through a 3.2mm opening in both the anode cover plate and extractor cone, across a 2.5mm gap. The cone tip is made out of molybdenum to reduce erosion due to extracted electrons that get bent due to the stray magnetic field of the source and strike the cone.

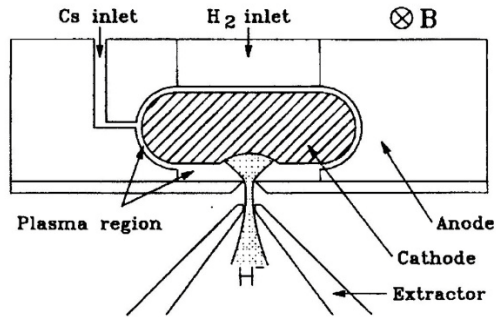

**FIGURE 4.** Round aperture magnetron schematic. (Image from reference [2])

The spherical focusing along with low arc current and high extraction voltage contribute to the power efficiency of the source which is 67mA/kW, as shown in Table 2. BNL experience has shown that the circular aperture ion source is very reliable, often running 6 months before needed maintenance.

**TABLE 2.** BNL operational parameters

| Parameter | Value |
|---|---|
| H- current | 90-100mA |
| Arc current | 8-18A |
| Arc Voltage | 140-160V |
| Extraction Voltage | 35kV |
| Rep Rate | 7.5Hz |
| Duty Factor | 0.5% |
| Pulse Width | 700µs |
| Power efficiency | 67mA/kW |

Ion source data taken to this point has been using a modified version of the HINS ion source [3] mounted on the source test stand. The source has been modified for a grounded extraction cone, and to match the BNL source geometry, apertures and cone spacing. The testing to this point has been a proof of principle before building the operational source and was not intended to optimize the source.

For current testing the rep rate is 7.5Hz, arc pulse width 300µs, and extraction pulse width of 200µs, which are very close to how BNL operates. With these parameters the maximum H- beam current witnessed so far in the test stand has been 70mA at 35kV extraction. Figure 5 shows a typical beam pulse measured on the toroid with 35kV extraction and arc current of 23A. The noise on the beam current flattop is about 20mA and is very dependent of the arc current. Higher arc currents tend to have less noise.

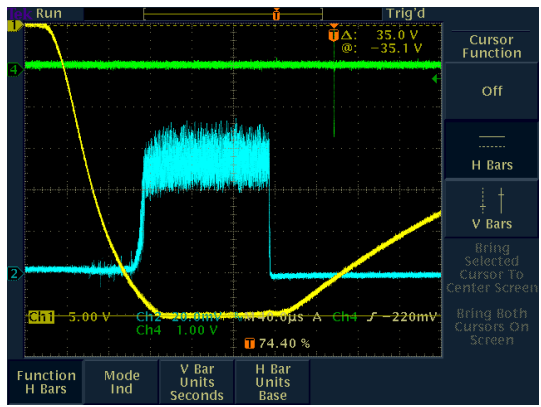

**FIGURE 5.** Typical scope trace of the source on the test stand operating at 35kV extrtaction.

For this project a new source vacuum chamber and source mounting are currently being designed and should be installed on the test stand within a couple of months. The design is a mixture of the HINS and BNL designs with ease of maintenance in mind. Figure 6 shows the preliminary layout of the vacuum chamber and source. The vacuum pumping will be 1200L/S for hydrogen and will be separated from the LEBT vacuum by a 3mm extraction aperture. This should allow for better vacuum in the beamline and reduce H- stripping losses.

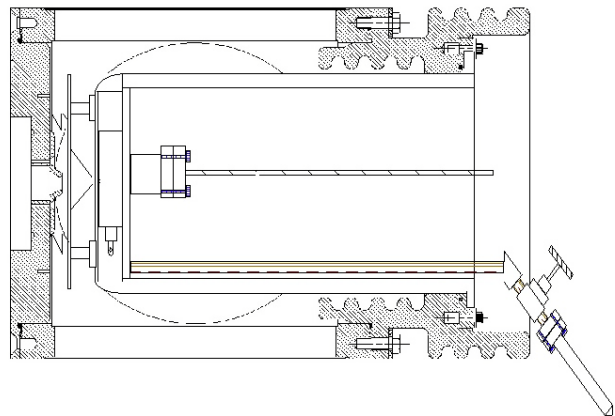

**FIGURE 6.** Proposed ion source vacuum chamber.

## Extraction

The source extraction shown in Figure 7 is different than the current operational system in that the extraction voltage is the acceleration voltage. This higher extraction voltage is more affective at pulling H- out of the source.

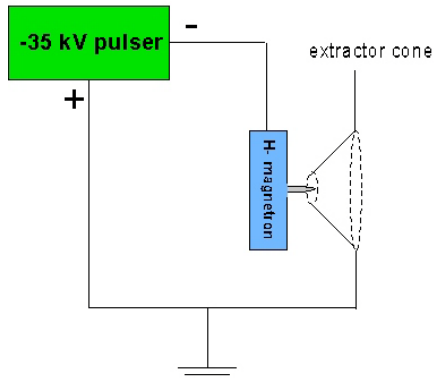

**FIGURE 7.** BNL extraction scheme. The extraction voltage is the accelerating voltage.

The negative 35kV extraction pulser design is a modified version of the FNAL extractor and similar to the one used at BNL. The pulser is capable of delivering 40kV, 400mA pulses at 15Hz. It pulses floating relay racks containing source electronics that are tied to the source body (anode) at -35kV. This provides the difference of potential for the extraction/accelerating voltage since the extractor is tied to ground.

## Einzel Lens Chopping

Chopping is needed to for changing the width of the beam pulse for different aspects of the physics program an Einzel lens located near the entrance to the RFQ will be used as a chopper for the leading edge of the beam [1]. As shown in Figure 8, the lens will initially be at 37kV which will stop the beam coming out of the source. The lens will then be set to 0V allowing the beam to pass through to the entrance of the RFQ. After the beam passes through the lens it be charged back up to 37kV prior to the next beam pulse.
.

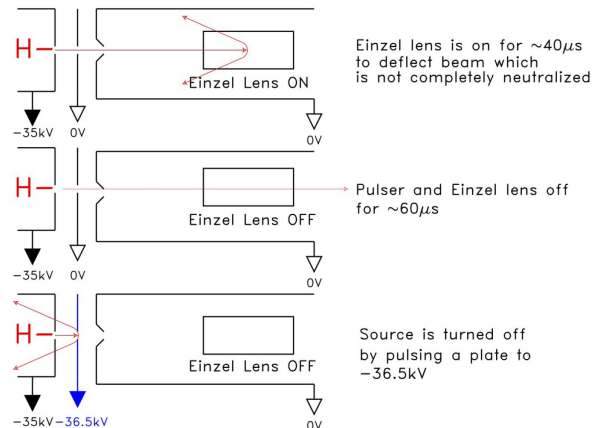

**FIGURE 8.** Proposed chopping using an Einzel lens. Figure is from reference [1].

There is an Einzel lens located just downstream of the ion source in the test stand that is used to focus the beam coming out of the source so that it will fit through a toroid and allow for emittance and Faraday cup measurements. This will allow measurement of the Einzel lens rise time which needs to be <1µs for chopping. Figure 9 shows SIMION simulations of the test stand with different voltages on the Einzel lens with a constant 35kV on the extractor.

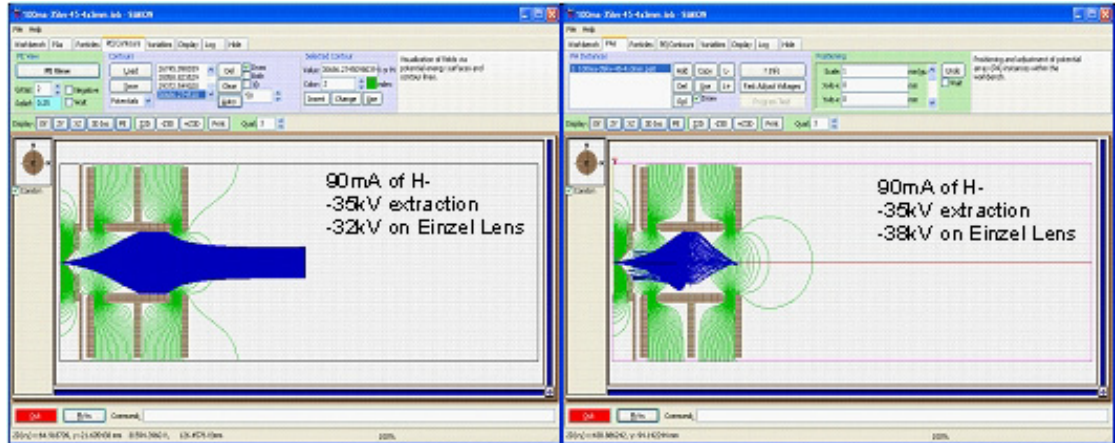

(a) SIMION simulation results for focused beam in the test stand.
(b) SIMION simulation for stopped beam in the test stand.

**FIGURE 9.** SIMION simulation results for Einzel lens test on the test stand.

For 35keV a lens voltage of 38kV will stop the beam coming out from making it through the lens. There is a pulser being built that will short the Einzel lens to ground with a thyratron tube and should have a rise time of ~50ns, which is much faster than what is need for chopping the beam.

## CONCLUSION

Before the new preinjector can be installed the magnetron ion source will need to be optimized. The source anode and extractor cone openings will need to have apertures that allow for enough H- to be extracted while maintaining the smallest emittance possible. Also, the extractor gap spacing will need to be large enough and the stray magnetic field optimized to sweep away the electrons (while not affecting the H- trajectories) to support at least 35kV extraction with a minimum amount of arcing.

# ACKNOWLEDGMENTS

I would like to thank Chuck Schmidt for all of his help and mentoring for this project. His decades of experience and willingness to help are always welcome. I would also like to thank the FNAL Preacc Group technicians for their help with the modifications to the HINS source.